# Population Pharmacokinetic Study of Tacrolimus in Pediatric Patients with Primary Nephrotic Syndrome: A Comparison of Linear and Nonlinear Michaelis–Menten Pharmacokinetic Model

Lingfei Huang[a, ‡], Yixi Liu[b,c ‡], Zheng Jiao[b*], Junyan Wang[a], Luo Fang[a], Jianhua Mao[d]

‡ Authors contributed equally

[a]Department of Pharmacy, National Clinical Research Center for Child Health, The Children's Hospital, Zhejiang University School of Medicine, Hangzhou, 310052, China
[b]Department of Pharmacy, Shanghai Chest Hospital, Shanghai Jiao Tong University, Shanghai, 200030, China
[c]Department of Pharmacy, Huashan Hospital, Fudan University, Shanghai, 200040, China
[d]Department of Nephrology, National Clinical Research Center for Child Health, The Children's Hospital, Zhejiang University School of Medicine, Hangzhou, 310052, China

[*]**Corresponding author**
Zheng Jiao, PhD
[1] Department of Pharmacy, Shanghai Chest Hospital, Shanghai Jiao Tong University,
West Huaihai Road 241,
Shanghai, China, 200030.
Tel: +86-21-5288 8712

[2] Department of Pharmacy, Huashan Hospital, Fudan University,
12 Urumqi Middle Road,
Shanghai, China, 200040.
Tel: +86-21-5288 8712

Email: jiaozhen@online.sh.cn

**Abstract**

*Background*

Little is known about the population pharmacokinetics (PPK) of tacrolimus (TAC) in pediatric primary nephrotic syndrome (PNS). This study aimed to compare the predictive performance between nonlinear and linear PK models and investigate the significant factors of TAC PK characteristics in pediatric PNS.

*Methods*

Data were obtained from 71 pediatric patients with PNS, along with 525 TAC trough concentrations at steady state. The demographic, medical, and treatment details were collected. Genetic polymorphisms of *CYP3A4*1G*, *CYP3A5*3*, and *ABCB1-C3435T* were analyzed. The PPK models were developed using nonlinear mixed-effects model (NONMEM®) software. Two modeling strategies, linear compartmental and nonlinear Michaelis–Menten (MM) models, were evaluated and compared.

*Results*

Body weight, age, daily dose of TAC, co-therapy drugs (including azole antifungal agents and diltiazem), and *CYP3A5*3* genotype were important factors in the final linear model (one-compartment model), whereas only body weight, co-therapy drugs, and *CYP3A5*3* genotype were the important factors in the nonlinear MM model. Apparent clearance and volume of distribution in the final linear model were 7.13 L/h and 142 L, respectively. The maximal dose rate ($V_{max}$) of the nonlinear MM model was 1.92 mg/day and the average concentration at steady state at half-$V_{max}$ ($K_m$) was 1.98 ng/mL. The nonlinear model described the data better than the linear model. Dosing regimens were proposed based on the nonlinear PK model.

*Conclusion*

Our findings demonstrate that the nonlinear MM model showed better predictive performance than the linear compartmental model, providing reliable support for optimizing TAC dosing and adjustment in children with PNS.

## 1 Introduction

Tacrolimus (TAC), a calcineurin inhibitor (CNI), has been widely used as second-line therapy for the pediatric primary nephrotic syndrome (PNS), which is manifested by massive proteinuria, hypoalbuminemia, edema, and hypercholesterolemia (Atanda, 2012; Lombel et al., 2013; Kaku et al., 2015; Jahan et al., 2015; Chinese Medical Association, 2017). TAC has low oral bioavailability and is transported out of the intestine, liver, and kidney cells via the efflux pump p-glycoprotein (P-gp) encoded by the *ABCB1*. It is metabolized by a mixed-function oxidase system, primarily comprising the cytochrome P-450 system (CYP3A), and it binds strongly to erythrocytes, albumin, and alpha-1-acid glycoprotein (Venkataramanan et al., 1995). TAC has a narrow therapeutic window with a wide inter-individual pharmacokinetic (PK) variability and therefore, close therapeutic drug monitoring (TDM) and dosage individualization are recommended.

Until recently, population pharmacokinetics (PPK) combined with the maximum a *posteriori* Bayesian (MAPB) estimation has been one of the classical methods to develop individualized dosing regimens (del Mar Fernández de Gatta et al., 1996; Fukudo et al., 2003; Woillard et al., 2017). The PPK of TAC has been extensively investigated in children after organ transplant, and the results have shown that body weight (Zhao et al., 2009; Musuamba et al., 2014; Prytuła et al., 2016; Andrews et al., 2018), *CYP3A5*3* genetic polymorphisms (Guy-Viterbo et al., 2014; Jacobo-Cabral et al., 2015; Prytuła et al., 2016; Andrews et al., 2018), and hematocrit (Zhao et al., 2009; Andrews et al., 2018) can partially explain the observed individual PK variability. However, so far, little is known about the TAC PK characteristics in children with PNS and the data in organ transplant patients might not be appropriate to extrapolate to patients with PNS, owing to the different pathological and physiological characteristics of PNS (Atanda, 2012). For example, gastrointestinal edema caused by hypoalbuminemia might affect the absorption of TAC and hypercoagulative state, one of the common complications of PNS, might change the distribution ratio of TAC in erythrocytes (Jahan et al., 2015).

To the best of our knowledge, only three studies (Huang et al., 2018; Hao et al., 2018; Wang et al., 2019) have investigated the PK of TAC in children with PNS. Our previous study (Huang et al., 2018) conducted in 100 children with PNS with 357 TAC trough concentrations showed that body weight, daily dose of TAC, and co-therapy with azole antifungal agents significantly influenced the clearance of TAC. In the other two studies, Wang et al. described a TAC PPK model based on age and body weight, whereas Hao et al. found the *CYP3A5*3* genotype as a covariate besides body weight. It is worth noting that, these studies used a linear compartmental model, a traditional PK modeling strategy for TAC; however, a dose-dependent clearance of TAC was found in our previous study (Huang et al., 2018). These findings contradict the traditional viewpoint that TAC is a drug with linear PK, as described by 1- or 2- compartment model (Musuamba et al., 2014; Jacobo-Cabral et al., 2015; Prytuła et al., 2016; Andrews et al., 2018; Hao et al., 2018; Wang et al., 2019), but is similar to a recent study in which a superior predictability of nonlinear Michaelis–Menten (MM) model for TAC was found in adult renal transplant recipients (Zhao et al., 2016).

We assumed that a nonlinear relationship might exist between dose and change in TAC steady-state PK. Disease status of pediatric PNS might result in nonlinear absorption or distribution process of TAC. For example, greater individual differences of oral bioavailability might occur when gastrointestinal edema exists in PNS. However, a linear model is inadequate to describe the absorption and distribution characteristic. Therefore, in this study, to the best of our knowledge, we adopted a nonlinear MM model for the first time to develop TAC PPK in pediatric PNS and then compared the results with those of the linear model. Furthermore, we included a higher number of patients sampled at more than three dose levels than our previous study, in order to improve the accuracy of judgment of non-linearity (Ahn et al., 2005).

Moreover, genetic polymorphism was not obtained in our previous study (Huang et al., 2018). The *CYP3A5\*3* genotype was identified as an important factor affecting TAC PPK in children after organ transplant (Guy-Viterbo et al., 2014; Jacobo-Cabral et al., 2015; Prytuła et al., 2016; Andrews et al., 2018), which was also demonstrated in a previous study (Hao et al., 2018) mentioned above in children with PNS. Smaller TAC doses are usually required for *CYP3A5\*3/\*3* carrier than for *CYP3A5\*1* carrier. In addition, due to the relatively high frequency of *CYP3A4\*1G* in Chinese subjects, studies (Shi et al., 2011; Zuo et al., 2013) have also indicated that a combination genotype of *CYP3A4\*1/\*1* and *CYP3A5\*3/\*3* might be clinically important for TAC dose design in adult organ transplant recipients and that it could improve the efficacy and tolerability of TAC. Besides, gastrointestinal edema of PNS might affect the activity of P-gp existing in the gastrointestinal tract, and thus, *ABCB1-C3435T* alleles, which has also been mentioned in some studies in adult organ transplant recipients (Chan et al., 2004; Wu et al., 2011; Kravljaca et al., 2016), was also investigated in this study in addition to *CYP3A5\*3* and *CYP3A4\*1G*.

Therefore, this study was conducted to compare the difference between the nonlinear and linear models based on routine TAC TDM data obtained from our center and to screen factors influencing TAC PPK characteristics in children with PNS, in order to develop individualized dosage recommendations based on the superior model.

## 2 Methods

### 2.1 Patients and clinic data collection

The concentrations of TAC in children with PNS were collected retrospectively from the medical records at the Children's Hospital, Zhejiang University School of Medicine, China, from April 2011 to July 2017. Patients with PNS younger than 18 years old stabilized on the same dose of TAC for ≥ 3 days were included. Patients with serious infections or multiple organ injuries were excluded. Patients co-administered with traditional Chinese medicines that might affect TAC concentrations were also excluded.

TAC was administered orally in the form of capsule (Prograf®, Astellas Ireland Co., Ltd. Killorglin, Co., Kerry, Ireland) every 12 hours. The initial dose of TAC was 50–150 μg/kg/day, which was later adjusted according to the routine TDM, clinical evidence of efficacy and adverse reactions. Complete records of actual body weight, height, age, gender, dosing history, sampling time, co-therapy drugs, and laboratory tests were also collected. The research had passed the ethical review of study hospital, and registered in Chinese Clinical Trial Registry (http://www.chictr.org.cn)(ID number: ChiCTR1900022943).

### 2.2 Determination of TAC concentration

Whole blood TAC concentrations were measured using an enzyme-multiplied immunoassay technique (Viva-E automatic biochemical analyzer, Siemens Healthcare Diagnostics Inc., Erlangen, Germany; EMIT® 2000 Tacrolimus Assay Kit, Siemens Healthcare Diagnostics Inc., Erlangen, Germany). Experimental procedures were conducted according to the manufacturers' instructions (Viva-E System, 2014). The calibration range was 2.0–30 ng/mL, and thus concentrations lower than 2.0 ng/mL were excluded. The inter- and intra-coefficients of variation for this assay were less than 10%.

### 2.3 Genotyping

Ethylenediaminetetraacetic acid (EDTA) anticoagulated blood was collected and delivered to Shanghai Xiangyin Biotechnology Company (Shanghai, China) for DNA extraction, polymerase chain reaction, and subsequent DNA sequencing (using the Snapshot technology, see **Appendix 1** for details). The genotyping of three single-nucleotide polymorphisms, namely, *CYP3A4*1G*, *CYP3A5*3*, and *ABCB1-C3435T*, were performed. The allele frequency was confirmed using Hardy-Weinberg equilibrium.

**2.4 Population pharmacokinetic modeling**

Data were analyzed using the nonlinear mixed-effects model program (NONMEM®, Version 7.4; Icon Inc, PA, USA) compiled with gFortran (Version 4.9.2; http://www.gfortran.org). The output was explored by the R package (Version 3.3.1; http://www.r-project.org) and Xpose (Version 4.5.3; http://xpose.sourceforge.net). The first-order conditional estimation method with η–ε interaction (FOCE-I) was used throughout the model-building procedure.

**2.4.1 Base model**

Two modeling strategies, linear PK modeling and nonlinear modeling (MM model), were employed in the model development separately. The linear PK model was a one-compartment (CMT) model with the first-order absorption and first-order elimination and was parameterized in terms of apparent total clearance (CL/F), absorption rate constant ($K_a$), and apparent volume of distribution ($V_d$/F). Because of no sampling in the absorption phase, the Ka was fixed to 4.48 h$^{-1}$ (Zhao et al., 2016).

The MM model investigated was shown as below:

$$\text{MM model: Dose} = \frac{V_{max} \times C_{ss}}{K_m + C_{ss}} \quad (1)$$

where Dose is oral dose of TAC. $C_{ss}$ represents observed concentration at steady-state after at least 6 oral doses. $V_{max}$ represents maximal daily dose at steady-state, and $K_m$ represents an MM constant equal to the steady-state trough concentration at half-maximum daily dose. The MM model was written in the $PRED block in NONMEM®.

Exponential models were used to account for between-subject variability (BSV) as below:

$$P_i = TV(P) \times \exp(\eta) \quad (2)$$

$\eta$ is defined as a symmetrically distributed, zero-mean random variable with a variance that is estimated as part of the model estimation.

The following statistical models describing the residual unexplained variability (RUV) were investigated (Eqs. 3-5):

$$Y = IPRED + \varepsilon \quad (3)$$
$$Y = IPRED \times \exp(\varepsilon) \quad (4)$$
$$Y = IPRED \times \exp(\varepsilon_1) + \varepsilon_2 \quad (5)$$

where Y represents the observation, IPRED is the individual prediction, and $\varepsilon_n$ represents the symmetrically distributed zero-mean random variables with variance terms that are estimated as part of the population model-fitting process.

**2.4.2 Covariate models**

The potential covariates included age, gender, body weight, height, hematocrit, albumin, aspartate transaminase, serum creatinine, estimated glomerular filtration rate [calculated using Schwartz formula (Schwartz and Work, 2009)], and co-therapy drugs, as well as genetic polymorphisms of *CYP3A4*1G*, *CYP3A5*3*, and *ABCB1-C3435T*. The daily dose of TAC was screened only for the linear model. Meanwhile, only the covariates with proportion > 5 % in all patients were investigated.

Body weight (WT) and age have significant impacts on the PK in pediatric patients. Therefore, physical maturation, which is a time-dependent process was considered at first. Four commonly used models based on allometric scaling were tested (Ding et al., 2015) (Eq. 6):

$$P_i = TV(P) \times \left(\frac{COV}{COV_{median}}\right)^\theta \times MF \qquad (6)$$

where $COV_{median}$ is the median of the covariate, MF is the maturation factor that is defined as the process of becoming mature. The model displaying the best fit was selected for further analysis.

*Model I*: In the simplest exponent model, the exponent θ was estimated, and MF was fixed to 1, indicating that maturation was not considered. This model is shown as Eq. 7:

$$P_i = TV(P) \times \left(\frac{COV}{COV_{median}}\right)^\theta \qquad (7)$$

*Model II*: For the maturation model, the exponent θ was assigned a fixed value of 0.75, and MF was calculated according to Eq. 8:

$$MF = \frac{1}{1+\left(\frac{Age}{TM_{50}}\right)^{Hill}} \qquad (8)$$

where $TM_{50}$ is the age at which clearance maturation reaches 50% of that of adults, and Hill is the slope parameter for the sigmoid $E_{max}$ maturation model.

*Model III*: This is referred to as the WT-dependent exponent model (Eq. 9):

$$\theta = \theta_0 - \frac{k_{max} \times WT^{Hill}}{k_{50}^{Hill} + WT^{Hill}} \qquad (9)$$

*Model IV*: This is referred to as the age-dependent exponent model (Eq. 10):

$$\theta = \theta_0 - \frac{k_{max} \times Age^{Hill}}{k_{50}^{Hill} + Age^{Hill}} \qquad (10)$$

where $\theta_0$ is the value of the exponent at a theoretical WT of zero (Eq. 9) or at birth (0 years) (Eq. 10), $k_{max}$ is the maximum decrease of the exponent, $k_{50}$ is the WT (Eq. 9) or age (Eq. 10) at which a 50% decrease relative to the maximum decrease is attained.

Akaike information criteria (AIC) and Bayesian information criteria (BIC) was employed in the selection of the competing non-nested aforementioned models, which was calculated using Pirana software (version 2.9.0, http://www.pirana-software.com/) (Keizer et al.,2011). Models with lower values of AIC and BIC were regarded as superior.

Continuous variables were investigated by linear models (Eqs. 11-12) or exponential model (Eq.13).

$$P_i = TV(P) + \theta \times \frac{COV}{COV_{median}} \qquad (11)$$

$$P_i = TV(P) + \theta \times (COV - COV_{median}) \qquad (12)$$

$$P_i = TV(P) \times \left(\frac{COV}{COV_{median}}\right)^\theta \qquad (13)$$

The categorical variables were investigated by scale model (Eq.14).

$$P_i = \begin{cases} TV(P) \times \theta & \text{if male} \\ TV(P) & \text{if female} \end{cases} \qquad (14)$$

where θ represents the degree of influence of covariate on the parameters, COV is value of a covariate for an individual and $COV_{median}$ is medium value or population typical value of covariate.

The covariate model was built in a stepwise manner, with forward inclusion and backward elimination. In the forward inclusion analysis, a decrease in the objective function value (OFV) of at least 3.84 ($\chi^2$ test, $P < 0.05$, *df*

= 1) was used as a criterion for inclusion of the covariate in the model. An increase in the OFV of at least 6.63 ($\chi^2$ test, $P < 0.01$, $df = 1$) was used as a criterion for retaining significant covariate-parameter relationships in the backward elimination step. The covariates in the model were also selected based on physiological plausibility of parameter estimates, goodness-of-fit plots, and statistical significance.

### 2.4.3 Model Evaluation

Prediction-based diagnostics and simulation-based diagnostics were used to compare the predictive performances of the linear model and non-linear model.

Goodness-of-fit (GOF) was used. GOF include observed concentrations (OBS) versus population predicted concentrations (PRED), OBS versus individual predicted concentration (IPRED), conditional weighted residuals (CWRES) versus PRED and CWRES versus time after dose.

Meanwhile, PRED and IPRED were estimated and compared with OBS by estimating the relative prediction error (PE%, Eq.15) and individual prediction error (IPE%, Eq.16), respectively. $IF_{20}$ and $IF_{30}$, which indicated the percentage of IPE that fell within the ± 20% and ± 30% range, respectively, were also applied to represent a combination index of both the accuracy and precision.

$$\text{PE (\%)} = \frac{\text{PRED} - \text{OBS}}{\text{OBS}} \times 100 \qquad (15)$$

$$\text{IPE (\%)} = \frac{\text{IPRED} - \text{OBS}}{\text{OBS}} \times 100 \qquad (16)$$

Nonparametric bootstrap (Ette et al., 1997) was employed to assess robustness of the model. 2000 bootstrap were performed using the Perl-speaks-NONMEM program (PsN, Version 4.8, http://psn.sourceforge.net). The median and the $2.5^{th}$-$97.5^{th}$ percentile intervals were calculated and compared with the values obtained using NONMEM®.

Normalized prediction distribution errors (NPDEs) were also investigated. The dataset was simulated 2000 times for NPDE, and the results were summarized graphically and statistically using the NPDE add-on package in R (Version 2.0, http://www.npde.biostat.fr/). NPDE results were summarized graphically using (1) quantile-quantile plot of the NPDE, (2) a histogram of the NPDE, (3) scatter plot of NPDE versus time, and (4) scatter plot of NPDE versus PRED. If the predictive performance is satisfied, the NPDE will follow a normal distribution (Shapiro-Wilk test) with a mean value of zero (t-test) and a variance of one (Fisher's test).

### 2.5 Dosing regimen optimization

Monte Carlo simulations were carried out using the parameter estimates from the final PPK model. The aim was to determine the optimal starting dosing regimen and achieve the target $C_{ss}$ of 5–10 ng/mL. Body weight of individual simulated was 10, 20, 30, 40, 50, 60, and 70 kg. One thousand simulations were carried out using the initial dataset and the $C_{ss}$ of each simulated subject was calculated.

Meanwhile, based on the actual bodyweight of patients for modelling, the TAC starting dose was calculated by two strategies: model-based estimation and empirical treatment in study hospital (50–150 μg/kg/day according to the recommendation of Chinese Medical Association). The predicted error of the model-based strategy was investigated (Eq.17):

$$\text{Predicted error (\%)} = \frac{D_{mbs} - D_{ets}}{D_{ets}} \times 100 \qquad (17)$$

where $D_{mbs}$ represents the starting dose of the model-based strategy, $D_{ets}$ represents the starting dose from empirical treatment strategy.

# 3 Results

## 3.1 Patient characteristics and gene allele frequency

In total, data of 71 children with PNS were collected in this study, and 525 whole blood TAC concentrations were available for modeling. Forty patients were sampled at two TAC dose levels at least, and the mean dose levels were 1.9 ± 0.9 per patient. The mean dose of TAC was 1.5 ± 0.6 mg/day and the corresponding mean TAC trough concentration was 5.6 ± 2.9 ng/mL. Patient demographics and laboratory tests used for model development are summarized in **Table 1**.

Genotype distribution and the allele frequency of *CYP3A4/5* and *ABCB1* are depicted in **Table 2**. There was no deviation from the Hardy-Weinberg equilibrium. The wild types of the *CYP3A4*1G* and *ABCB1-C3435T* genes were 64.8% and 33.8%, respectively. Patients with *CYP3A5*1/*1*, the wild type, was not be involved due to the low proportion (<5 %) in all patients; therefore, the covariate effect of *CYP3A5*3* was focused on *CYP3A5*1*3* and *CYP3A5*1*3* genotypes.

## 3.2 Model Building

### 3.2.1 Linear PK model

A one-compartment (CMT) model with first-order absorption and elimination was employed to describe the linear PK of TAC. A residual model with the exponential method was chosen and the population typical value of CL/F and $V_d$/F in the base CMT model was 8.56 L/h and 202 L, respectively. In the forward inclusion step, by exploring the relationship of CL/F with body weight and age, the simplest allometric model (*Model I*) had lower AIC and showed the best fit. The daily dose of TAC and the combination of azole antifungal agents, diltiazem, and amlodipine were shown to have significant effects on the CL/F. In the backward elimination step, all the covariates remained in the final model except co-therapy with amlodipine.

The CL/F and $V_d$/F in the final CMT model was 7.13 L/h and 142 L, respectively, for a 25-kg and 2-years old PNS pediatric patient with *CYP3A5*3/*3* allele administered with 1.5 mg/day. The final model for CL/*F* was listed as below:

$$\text{CL/F (L/h)} = 7.13 \times \left(\frac{DD}{1.5}\right)^{0.225} \times \left(\frac{WT}{25}\right)^{0.265}$$

$$\times 1.394 \text{ (if } CYP3A5*1*3\text{)}$$

$$\times 0.538 \text{ (if combined with azole antifungal agents)}$$

$$\times 0.88 \text{ (if combined with diltiazem)} \quad (18)$$

where WT is the bodyweight and DD is the daily dose of TAC.

It is noteworthy that when the daily dose of TAC was incorporated, the predictive performance of the final model obviously improved. A nonlinear positive correlation between the daily doses and CL/F showed that the higher the daily dose, the higher the CL/F.

### 3.2.2 Michaelis–Menten model

A combined model based on additive and exponential residual methods was chosen, and the population typical value of $V_{max}$ and $K_m$ in the base MM model were 1.8 mg·day$^{-1}$ and 1.04 ng/mL, respectively. The correlations between $V_{max}$ and $K_m$ was considered, and the impact of all covariates were tested in $V_{max}$ and $K_m$ separately. Body weight was an important covariate for both $V_{max}$ and $K_m$, and the simplest allometric model (*Model I*) was used. *CYP3A5*3*3* genotype, co-therapy with azole antifungal agents and diltiazem also affected $K_m$ after the stepwise screening. The $V_{max}$ and $K_m$ in the final MM model were 1.92 mg/day and 1.98 ng/mL, respectively. The final MM model was listed as below:

$$V_{max} \text{ (mg/day)} = 1.92 \times \left(\frac{WT}{25}\right)^{0.559} \tag{19}$$

$$K_m \text{ (ng/mL)} = 1.98 \times \left(\frac{WT}{25}\right)^{0.559}$$
$$\times 0.189 \text{ (if } CYP3A5*1*3\text{)}$$
$$\times 4.12 \text{ (if combined with azole antifungal agents)}$$
$$\times 2.01 \text{ (if combined with diltiazem)} \tag{20}$$

### 3.3 Model evaluation

The GOF plots indicated sufficient fit to the final models presented in **Fig. 1**. The final models showed no obvious bias or significant trends within these scatter plots, and the data fitting was considerably improved relative to that of the base model. Furthermore, the predictive performance of the models is presented as PE and IPE in the box plot (**Fig. 2**). The $IF_{20}$ (85.5%) and $IF_{30}$ (92.4%) of MM models reached > 80%, implying its superiority compared with < 70% of the CMT model ($IF_{20}$ and $IF_{30}$ was 47.2% and 68.4%, respectively).

**Table 3** shows the summary of the estimates obtained using the final CMT and MM models from bootstrap data. The two final models were stable, and the median values from the bootstrap procedure were close to the parameter estimates from the NONMEM, with < 5% bias. Meanwhile, the success rate determined using the bootstrap method of the MM model was 99.0%, higher than that of the CMT model (81.0%).

The NPDE results are presented in **Fig. 3**. The assumption of a normal distribution for the differences between predictions and observations was acceptable. The quantile-quantile plots and histogram also confirmed the normality of the NPDE, indicating good predictability of the final MM model.

### 3.4 Dosing regimens

Due to the better predictive performance of the nonlinear MM model, a dosing algorithm (**Table 4**) for the starting dose of TAC was calculated based on the final established MM model:

$$\text{Dose} = \frac{V_{max \text{ (final)}} \times C_{ss}}{K_{m \text{ (final)}} + C_{ss}} \tag{21}$$

where $V_{max(final)}$ represents Eq.19, and $K_{m(final)}$ represents Eq.20.

The required dose can be calculated using the $V_{max}$, $K_m$, and the desired target trough concentration. For example, a *CYP3A5*3*3* expresser weighing 20 kg without co-therapy of azole antifungal agent and diltiazem, can achieve a target steady-state trough concentration of 5–10 ng/mL, when the starting TAC dose ranged from 1.26–1.47 mg/day.

Meanwhile, **Fig. 4** shows the predicted error of the TAC starting dose computed by MM model when compared to the empirical strategy. The mean value of predicted error was 10.8% with standard deviation of 12.23%, which indicated a stable predictability of proposed strategy in MM model.

## 4 Discussion

TAC was considered to have linear kinetics in most of the previous studies. In this study, we also established a one-compartment linear PK model based on bodyweight, daily dose of TAC, *CYP3A5*3* genotype, and co-therapy drugs. Interestingly, the incorporation of daily dose improved the predictive performance of the final model significantly. Consequently, our study aimed to explore a specific answer to the question of whether TAC steady-state pharmacokinetics change with oral dose rate; therefore, a non-linear behavior of TAC PK in pediatric PNS patients was first determined and compared to the linear model.

To the best of our knowledge, the MM model was an effective approach to describe the non-linear relationship of dose and steady state trough concentration. MM models in our study are data-driven models that are not based on mammal compartment models. It is noteworthy that the MM model is a description of overall non-linearity, could one or several processes in PK, including absorption, distribution, metabolism, and excretion *in vivo*. Our finding is consistent with previous population analysis of tacrolimus in renal transplant patients (Zhao et al., 2016) and cyclosporine, another commonly used calcineurin inhibitor in renal transplant patients (Greve et al., 1993; Mao et al., 2018).

In our present MM model, the typical value of $V_{max}$ in pediatric PNS was 1.92 mg/day (95% CI, 1.76-2.19 mg/day), which indicated that the dose of TAC in excess of 1.92mg/day was likely to increase the concentrations unexpectedly. The typical value of $K_m$ in pediatric PNS was 1.98 ng/mL (95% CI, 0.77–2.11 ng/mL); that is, the nonlinearity of TAC will no longer influence dose rate adjustments when the $C_{ss}$ was at or below 1.98 ng/mL. However, the values of $V_{max}$ and $K_m$ in pediatric PNS were significantly lower than those in adult renal transplant recipients ($V_{max}$ was 5.54 mg/day and $K_m$ was 2.36 ng/mL) (Zhao et al., 2016). The discrepancy in these two populations might be due to various changes during the growth of children (Yokoi, 2009) and higher target TAC trough concentrations in renal transplant patients [10–15, 8–15, and 5–12 ng/mL within 1, 1–3, and 3–12 months after transplantation, respectively (KDIGO Work Group, 2009)] than that in PNS patients [5–10 ng/mL] (Atanda, 2012; Chinese Medical Association, 2017).

We hypothesized that an important factor that contributes to the nonlinear process of TAC was the large absorption variability, a typical feature of TAC caused by its high lipophilicity and transportation out of the intestine via P-gp. Furthermore, when the gastrointestinal environment changes (for example, gastrointestinal edema in PNS or gastrointestinal dysfunction post organ transplantation), higher fluctuation in absorption saturation became more pronounced.

The second possible explanation for the nonlinear PK of TAC might be the drug distribution complexity in children with PNS. Descending plasma protein binding rate caused by hypoalbuminemia of PNS might increase the proportion of free TAC; on the contrary, hypercoagulability, a common complication of PNS, might lead to higher TAC binding rate to erythrocytes, and then decrease the proportion of free TAC (Zahir et al.,2004; Golubović et al., 2014). However, the relationship between nonlinear PK and distribution *in vivo* could not be explained by the present study results, as albumin and hematocrit were not shown as significant variables in the established models.

Moreover, the immature hepatic enzyme metabolism capability, relative liver weight, hepatic blood flow, and nutrition in children especially in infants younger than three years of age, might also attribute to the nonlinear metabolic process of TAC (Yokoi, 2009).

In general, this interesting finding of MM model provides a novel perspective for future investigations; however,

considering that only trough samples were available, the function of daily dose might mostly reflect the non-linearity of the clearance. Thus, further studies are needed to confirm the above assumptions and to explore the physiological significance of $V_{max}$ or $K_m$.

Both the linear and nonlinear models indicated that WT could significantly affect the PK of TAC in pediatric with PNS. Several structural models based on WT were tested in this study, and the maturation model best fit the data in the CMT model, whereas the allometric model for MM model. We also found that there was a positive correlation between WT and CL/F or $V_{max}$; therefore, a higher WT-normalized TAC dose was required as WT increased.

The *CYP3A5*3* genotype was identified as the most influential factor in all tested genetic polymorphisms in our study, in agreement with previously published data in pediatric and adult organ transplant patients (Shi et al., 2011; Zuo et al., 2013; Guy-Viterbo et al., 2014; Jacobo-Cabral et al., 2015; Prytuła et al., 2016; Andrews et al., 2018). Patients carrying *CYP3A5*1*3* and *CYP3A5*1*1* had a higher CL/F or $V_{max}$ than *CYP3A5*3/*3* carriers with non-functional drug metabolic protein. Thus, *CYP3A5*3* genotyping might be a useful approach for better prediction of individual TAC doses in pediatric PNS. For the low frequency of *CYP3A5*1/*1* genotype among Chinese population (approximately 2.0–9.3% in previous reports (Shi et al., 2011; Zuo et al., 2013) and 2.7% in our study), a lower initial TAC dose should be recommended for patients in whom the CYP3A5 genotype is unknown. Dose recommendations in **Table 4** might serve as a reference for TAC initial dose in children with PNS. In addition, the *CYP3A4*1G* and *ABCB1 C3435T* alleles, which were mentioned in some studies in adult organ transplant recipients (Wu et al., 2011; Shi et al., 2011; Zuo et al., 2013; Kravljaca et al., 2016), were not considered as important covariates in either linear or nonlinear model, as the incorporation of these two alleles could not improve the predictive performance of the final models.

Co-medications also contributed to the difference in TAC PK in pediatric PNS. Azole antifungal agents, such as ketoconazole (Khan et al.,2014) and fluconazole (Guy-Viterbo et al., 2014) could inhibit the activity of intestinal and hepatic CYP3A enzymes and thus reduce TAC clearance, leading to a higher level of TAC concentration. Similarly, diltiazem, a benzothiazepine calcium channel blocker (CCB), is frequently used as a TAC-sparing agent (Jones and Morris, 2002; Li et al., 2011). However, amlodipine, another CCB used in our center for treating hypertension, was not considered to be a covariate in our final models. Several researchers (Passey et al., 2011; Storset et al., 2014; Lunde et al.,2014) regarded corticosteroids as an important factor; however, there was no difference with or without steroids in our study, probably because of the low dose of concomitant steroids (13.1 ± 10.7 mg/day presented as prednisone).

There are some limitations to our study. First of all, for the retrospective property of this study, only trough concentrations were obtained; therefore, only CL/F and its covariate could be reliably assessed in the current study (Booth et al., 2003; Kowalski et al., 2001). Trough concentrations alone were not adequate to estimate Vd, which was also shown by the little between-subject variability estimation on Vd. For the same reason, it is hard to clear up what is the specific non-linear PK process. Secondly, the allele frequency of *CYP3A5*1*1*, the wild type, was very low in Chinese population, and high RSE was found when fixed the parameter of *CYP3A5 *1* allele (*CYP3A5 *1*3/ CYP3A5 *3*3*) as a covariate in the preliminary analysis of this study. Thirdly, parameter estimates for *CYP3A5*1*3* and co-therapy of diltiazem showed high precision RSE, which may be attributable to insufficient information from the small patient size and sparse samples. Further investigation will be needed.

**5 Conclusion**

TAC PK in children with PNS was first compared between the linear compartment and nonlinear models, and factors affecting individual PK variability were identified. MM model, the nonlinear model, described the PK behavior of TAC better than the linear model. Bodyweight, *CYP3A5*3* genotypes, and co-therapy with azole antifungal agents and diltiazem were found to be significant in the final MM model, and dosing regimens were proposed. Further investigations are required to explore the nonlinearity of TAC.

**Table 1** Demographic data of pediatric PNS patients

| Characteristic | Mean ± SD | Median (Range) |
|---|---|---|
| Number of patients (Male/Female) | 71 (51/20) | / |
| Number of TAC samplings | 525 | / |
| Age (years) | 7.8 ± 3.7 | 7.3 (1.7-16.1) |
| Body weight (kg) | 28.4 ± 14.1 | 25.0 (10.0-83.0) |
| Height (cm) | 121.4 ± 23.1 | 118.0 (80.0-170.0) |
| TAC daily dose (μg/kg/day) | 59.8 ± 23.3 | 57.1 (6.4-160.0) |
| TAC daily dose (mg/day) | 1.5 ± 0.6 | 1.5 (0.25-4.0) |
| TAC concentration (ng/mL) | 5.6 ± 2.9 | 5.0 (2.0-29.2) |
| TAC dose level per patient | 1.9 ± 0.9 | 2.0 (1.0-4.0) |
| Haematocrit (%) | 41.6 ± 4.1 | 41.8 (19.1-52.4) |
| Albumin (g/L) | 27.8 ± 11.0 | 30.6 (7.9-54.6) |
| Aspartate transaminase (U/L) | 41.5 ± 21.0 | 37.0 (8.0-139.0) |
| Serum creatinine (μmol/L) | 40.3 ± 16.2 | 38.5 (4.0-129.0) |
| eGFR (mL/min/1.73 m$^2$) | 175.5 ± 90.5 | 154.1 (49.4-398.2) |
| Co-therapy medications | | |
|   Corticosteroid [a] | 71 (481) | / |
|     Corticosteroid daily dose (mg/day) [b] | 14.5 ± 11.0 | 10.0 (0.0-75.0) |
|   Calcium channel blockers | | |
|     Diltiazem [a] | 17 (111) | / |
|     Amlodipine [a] | 9 (38) | / |
|   Azole antifungal agent [a] | 4 (15) | / |

*TAC,* tacrolimus; *eGFR,* estimated glomerular filtration rate, calculated based on serum creatinine using the Schwartz formula as follow: eGFR = k × Height (cm) / Serum creatinine (μmol·L$^{-1}$), where k is a constant (for 1-12 years boys and >1 year girls, k is 49; and for >12 years boys, k is 62).

[a] Presented as number of patients (samples)
[b] Total daily prednisone dose, with methylprednisolone dose immerged after standardization

**Table 2** Allele frequencies of genetic polymorphisms in *CYP3A4*1G*, *CYP3A5*3* and *ABCB1-C3435T* genes

| Genotypes | Number of patients | Percentage |
|---|---|---|
| *CYP3A4*1G* (G20230A, rs2242480) | | |
| GG (*1/*1) | 46 | 64.8 % |
| GA (*1/*1G) | 21 | 29.6 % |
| AA (*1G/*1G) | 4 | 5.6 % |
| *CYP3A5*3* (A6986G, rs776746) | | |
| GA (*1/*3) | 26 | 36.6 % |
| GG (*3/*3) | 45 | 63.4 % |
| *ABCB1-C3435T* (rs1045642) | | |
| CC | 24 | 33.8 % |
| CT | 38 | 53.5 % |
| TT | 9 | 12.7 % |

All frequencies were in agreement with those predicted by Hardy-Weinberg equation ($P > 0.05$)

**Table 3** Population pharmacokinetic parameter estimates of the base model, final model, and bootstrap

| Parameter | Base model | | Final model | | Bootstrap of final model | | Bias (%) |
|---|---|---|---|---|---|---|---|
| | Estimate | RSE (%) | Estimate | RSE (%) | Median | 2.5%-97.5% | |
| **Linear one compartment model** | | | | | | | |
| OFV | 1415.210 | | 1265.676 | | | | |
| $k_a$ (h$^{-1}$) | 4.48 | / | 4.48 | / | 4.48 | / | / |
| CL/F (L/h) | 8.56 | 6.6 | 7.13 | 6.5 | 7.08 | 4.66-8.00 | -0.70 |
| *CYP3A5*1*3* | / | / | 0.394 | 24.1 | 0.393 | 0.209-0.620 | -0.25 |
| Body weight | / | / | 0.265 | 32.3 | 0.267 | 0.122-0.481 | 0.75 |
| Daily dose of tacrolimus | / | / | 0.225 | 34.8 | 0.213 | 0.028-0.335 | -5.33 |
| Co-therapy with Azole antifungal agent | / | / | -0.462 | 6.1 | -0.458 | -0.548-(-0.212) | -0.87 |
| Co-therapy with diltiazem | / | / | -0.12 | 40.9 | -0.125 | -0.219-(-0.014) | 4.17 |
| $V_d/F$ (L) | 202 | 23.7 | | | | | |
| Between-subject variability | | | | | | | |
| CL/F (%) | 38.9 | 27.6 | 23.9 | 25.5 | 23.4 | 15.8-30.3 | -2.09 |
| Residual variability | | | | | | | |
| Proportional error (%) | 36.1 | 10.3 | 32.9 | 9.5 | 32.4 | 29.3-35.7 | -1.52 |
| **Nonlinear Michaelis-Menten model** | | | | | | | |
| OFV | -661.967 | | -813.585 | | | | |
| $V_{max}$ (mg/day) | 1.8 | 5.7 | 1.92 | 4.4 | 1.92 | 1.76-2.19 | 0.00 |
| Body weight | / | / | 0.559 | 15 | 0.553 | 0.255-0.671 | -0.51 |
| $K_m$ (ng/mL) | 1.04 | 26.3 | 1.98 | 17.9 | 1.97 | 0.77-2.11 | 0.00 |
| *CYP3A5*1*3* | / | / | 0.189 | 38.1 | 0.183 | 1.42-9.78 | -0.99 |
| Body weight | / | / | 0.559 | 15 | 0.553 | 0.255-0.671 | 0.00 |
| Azole antifungal agent | / | / | 3.12 | 9.7 | 3.12 | (-0.55)-0.018 | -1.07 |
| Diltiazem | / | / | 1.01 | 22.1 | 1.00 | 0.352-1.78 | -0.51 |
| Between-subject variability | / | / | | | | | |
| $V_{max}$ (%) | 35.2 | 19.7 | 20.4 | 21.2 | 20.0 | 14.8-24.3 | -1.96 |
| $K_m$ (%) | 93.0 | 35.8 | 58.6 | 32.4 | 56.8 | 38.1-82.9 | -3.07 |
| Residual variability | | | | | | | |
| Proportional error (%) | 17.6 | 15.3 | 16.1 | 14.7 | 16.0 | 13.6-18.1 | -0.62 |

***OFV*,** Objective function value; ***RSE,*** relative standard error; ***Bias%*** = (Bootstrap – NONMEM) / NONMEM × 100%

**Table 4** Dosing regimens based on non-linear model for the starting dose of tacrolimus with a target steady-state trough concentration ($C_{ss}$) of 5 ng/mL and 10 ng/mL.

| Weight (kg) | $C_{ss}$ (ng/mL) | Dose (mg/day) | | | |
|---|---|---|---|---|---|
| | | *CYP3A5*1*3* expresser | | *CYP3A5*3*3* expresser | |
| | | without co-medications | with co-medications | without co-medications | with co-medications |
| 10 | 5 | 1.10 | 0.84 | 0.93 | 0.39 |
| | 10 | 1.15 | 0.97 | 1.05 | 0.59 |
| 20 | 5 | 1.59 | 1.10 | 1.26 | 0.44 |
| | 10 | 1.67 | 1.33 | 1.47 | 0.71 |
| 30 | 5 | 1.96 | 1.26 | 1.48 | 0.46 |
| | 10 | 2.08 | 1.58 | 1.78 | 0.77 |
| 40 | 5 | 2.28 | 1.38 | 1.65 | 0.47 |
| | 10 | 2.43 | 1.78 | 2.03 | 0.81 |
| 50 | 5 | 2.55 | 1.48 | 1.79 | 0.49 |
| | 10 | 2.73 | 1.94 | 2.24 | 0.85 |
| 60 | 5 | 2.79 | 1.56 | 1.90 | 0.50 |
| | 10 | 3.01 | 2.08 | 2.42 | 0.87 |
| 70 | 5 | 3.01 | 1.62 | 2.00 | 0.51 |
| | 10 | 3.26 | 2.20 | 2.58 | 0.89 |

*co-medications,* azole antifungal agent and diltiazem

### (a) Compartment model

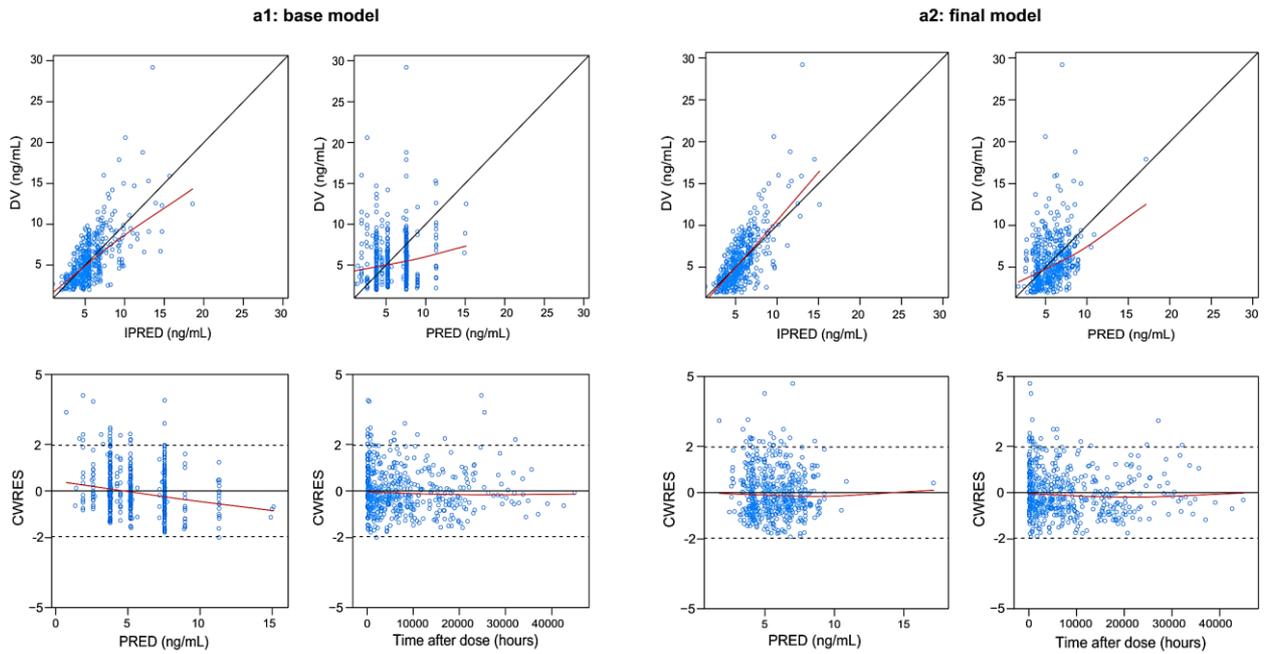

### (b) Michaelis-Menten model

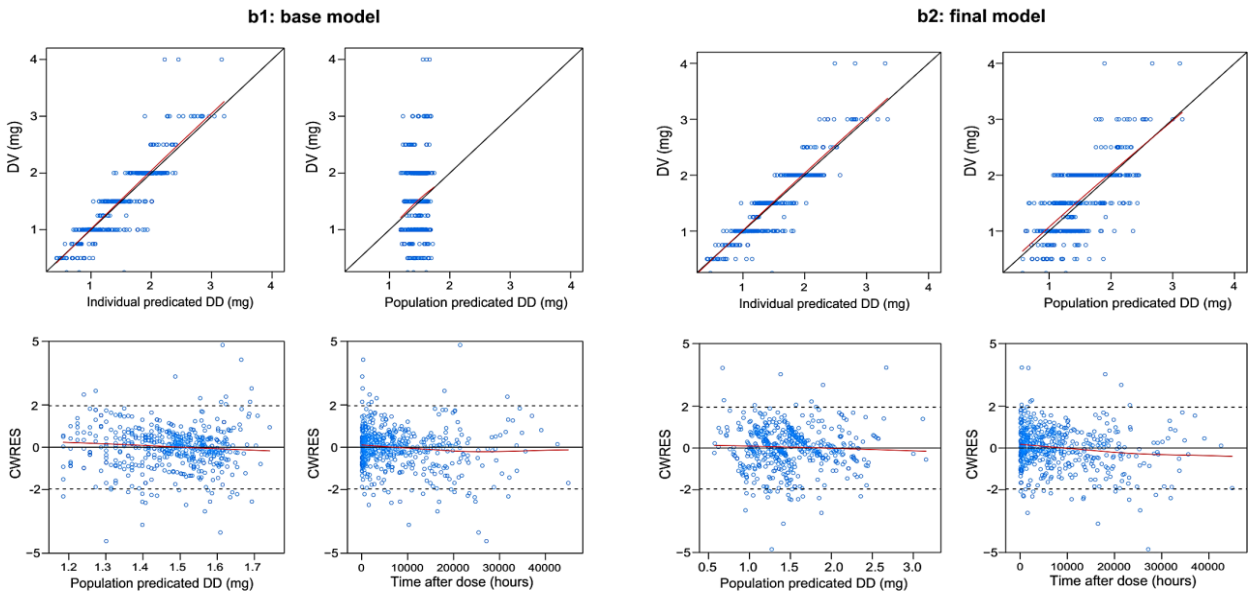

**Fig. 1**

Goodness-of-fit plot for the **(a)** compartment model and **(b)** Michaelis-Menten model. ***DV,*** dependent variable; ***IPRED,*** individual prediction; ***PRED,*** population prediction; ***CWRES***, conditional weighted residual; ***DD***, daily dose.

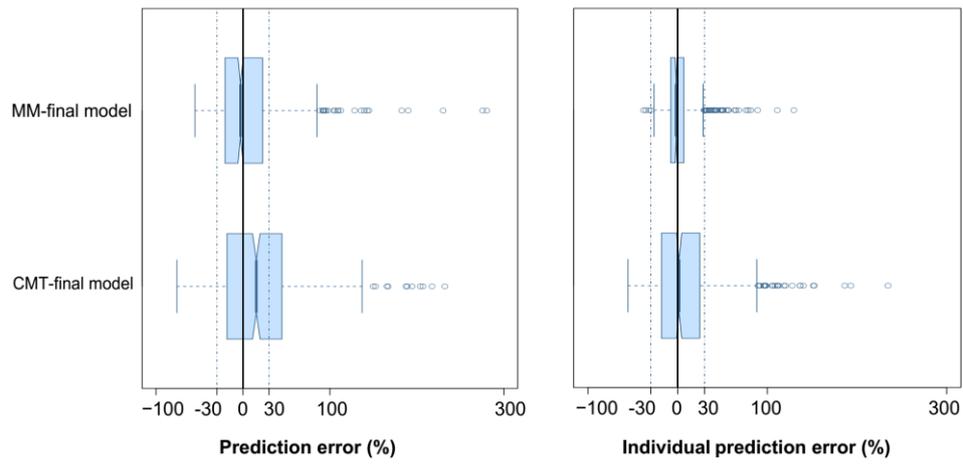

**Fig. 2**

Box plots of prediction error (%) and individual prediction error (%) of Michaelis-Menten (**MM**) final model and compartment (**CMT**) final model.

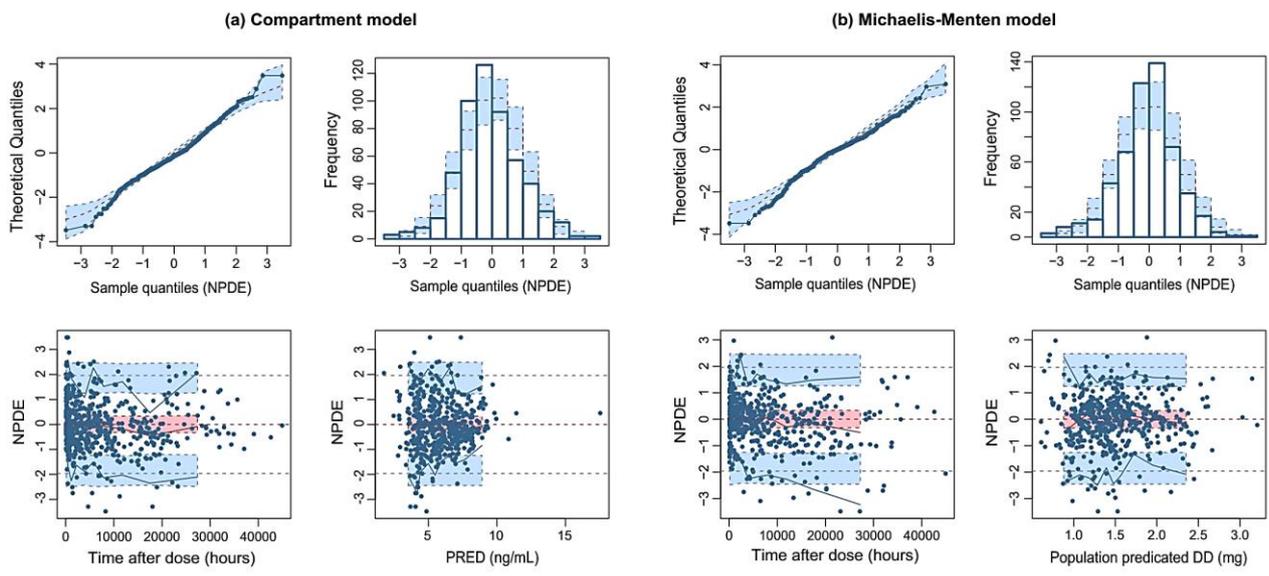

**Fig. 3**

NPDE of the two final models: **(a)** compartment model and **(b)** Michaelis-Menten model. *NPDE*, normalized predictive distribution error; *PRED*, population prediction; *DD*, daily dose.

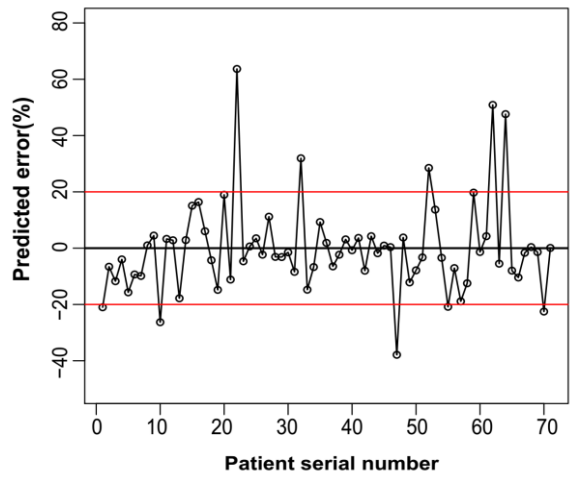

**Fig. 4**

Predicted error of TAC starting dose estimated based on the final Michaelis−Menten model.